\documentclass[preprint,authoryear]{elsarticle}

\usepackage{bm}
\usepackage{color}

\begin{document}


\title{
{Generalized Faraday law derived from classical forces in a rotating frame }}

\author[rvt]{Taeseung Choi\corref{cor1} }
\ead{tschoi@swu.ac.kr}

\address{ Barom Liberal Arts College, Seoul Women's University, Seoul 139-774, Korea} 

\cortext[cor1]{Corresponding author}
\fntext[fn1]{fax: +82 2 9705907.}

\date{\today}

\begin{abstract}
We show the additional spin dependent classical force due to the rotation of an electron spin's rest frame
is essential to derive a spin-Faraday law by using an analogy with the usual Faraday law. 
The contribution of the additional spin dependent force to the spin-Faraday law is
the same as that of the spin geometric phase. 
With this observations, Faraday law is generalized to       
 include both the usual Faraday and the spin-Faraday 
laws in a unified manner.
\end{abstract}


\maketitle

\section{Introduction}

Recently generalizations of Faraday law which includes a spin geometric phase 
or Berry phase \cite{Berry1} have been studied 
and these motive forces are measured also \cite{Barnes, Yang1}. 
The spin motive force is induced when the spin vector potential 
defined by the spin geometric phase is time varying. 
The generalization of the motive force to include the spin Berry phase
was obtained by using the mathematical equivalence of the Berry phases of
electromagnetic and spin origin \cite{Griffiths}.
This mathematical equivalence, however, has still a lack of a physical origin.
In this paper we have studied the physical origin of the generalized motive force using
a classical theory of a spin magnetic moment. 

The Aharonov-Bohm (AB) phase due to a magnetic vector potential is a manifestation
of the Berry phase \cite{AB}. In the original work Aharonov and Bohm have
introduced a scalar counterpart (SAB) caused by a scalar electric potential in the 
Schr\"odinger equation. 
In 1984 Aharonov-Casher discovered a phase acquired by a neutral particle with
a magnetic moment encircling a line of charge has been as the "dual" 
of the vector AB phase \cite{AC} and 
has been experimentally verified for thermal neutron \cite{Cimmino}
and for an atomic system \cite{Sangster}.
There is also a scalar AC (SAC) phase accumulated by a neutral particle in magnetic fields
during cyclic motion though experiencing no force \cite{Anandan}. (In some literature,
this is referred to as SAB phase \cite{Badurek}.) 
In a fundamental generalization of Berry's idea, Aharonov and Anandan (AA) has lifted 
the adiabatic restriction and could define a nonadiabatic geometric phase, an AA phase
\cite{AA}. 


The AC effect has been extended to include electronic systems
\cite{Mathur} and the extension for SAC effect is parallel. 
In AC and SAC effects spin-orbit and Zeeman interactions can be
interpreted as non-Abelian vector and scalar potentials coupled to the electronic spin
respectively \cite{Goldhaber, Anandan}. 
We call these non-Abelian gauge fields as $SU(2)_{spin}$ gauge fields since
the non-Abelian gauge structure stems from the interaction between electromagnetic fields 
and electronic spins.


{When the AB phase is time varying, 
an electromotive force is induced by Faraday's law
of magnetic induction, $\mathcal{E} = -\frac{1}{c}\frac{d}{ dt}\Phi_B
=-\frac{1}{2\pi}\frac{d}{dt}\left(\Phi_{AB} \Phi_0 \right)$
\cite{Jackson}, where $\Phi_B$ is a magnetic flux surrounded by a closed path and 
$\Phi_{AB}$ and $\Phi_0={hc}/{e}$ are the AB phase and
one flux quantum respectively.  
The SAB phase does not contribute to the motive force since
the electric potential $\phi$ that could generate the SAB phase 
is associated with a conservative electric field $-\nabla \phi$.}

The SAC phase, however, contributes to a motive force when it becomes
time-dependent.
Stern has noticed that the time-dependent Berry phase accumulated by the
electronic spin encircling a ring under the Zeeman interaction induces a
motive force motivated by the similarity between the Berry phase and
the AB phase \cite{Stern}.
Aronov and Lyanda-Geller pointed out that the
time-dependent AC flux due to spin-orbit interactions induces a motive
force \cite{Aronov1}. Balatsky and Altshuler argued that a spin
motive force can be induced via the Faraday law similarly to Stern's
motive force \cite{Balatsky}.

According to the dual nature between AB and AC effects 
the similar Faraday law such as 
 $\mathcal{E}_s = -\frac{1}{2\pi}\frac{d}{dt} \left(\Phi_{AC}\Phi_0\right)$ 
 is expected to be satisfied  for the magnetic moment, 
 when the AC phase $\Phi_{AC}$ is time-varying.
 This could be referred to as spin-Faraday law in analogy with usual Faraday law.
 Ryu \cite{Ryu} has given a unified view for various spin-motive
forces and spin-Faraday laws using a gauge theoretic approach.
He has concluded, however, that the exact parallelism may not exist between
the spin-motive force and the electromotive force, that is,
the spin-Faraday law associated with the AC flux is sometimes not valid.
He argued this by obtaining a covariant force on a spin from the classical 
Lagrangian based on the $SU(2)_{spin}$ gauge theory. 


 We will show in this paper that a spin dependent force newly added due to 
 the rotation of a rest frame of the electron spin is essential
 to derive the spin-Faraday law by using an analogy with the usual Faraday law.
 In the rotating frame the spin direction is set equal to
 the $z$-axis of the rest frame so that the additional spin dependent force depends
 on the electron spin precession. As a result, the line integral of this force
for a cyclic path is proportional to the solid angle subtended by the closed trace
of the electron spin. Therefore the contribution of this spin dependent force to
the spin-Faraday law is the same as that of the spin geometric phase which is half the solid
angle subtended by the spin precession.
In section \ref{sec:AC} we will re derive the AC phase for self-containedness
using invariant operator method, which is useful to solve the Schr\"odinger-type equations. 
In section \ref{sec:CF} we will discuss the classical theory for an electron with spin interactions
in the rotating rest frame of the electron spin and show the spin-Faraday law is valid
in the AC case. 
The extension of Stern's result to nonadiabatic case and its observability are discussed.
With this observations, a Faraday law is generalized to       
 include both the usual Faraday and the spin-Faraday 
laws in a unified manner. In section \ref{sec:DIS} we summarize and discuss our results.
 
\section{Model and AC phase revisited}
\label{sec:AC}
The nonrelativistic Hamiltonian for an electron in external 
electromagnetic potentials reduced from the Dirac Hamiltonian in the low-energy field limit 
is written by 
\begin{equation}
\label{Eq:One}
\mathcal{ H }=  \frac{1}{2m} \left( {\bf{p}} - \frac{e}{c} {\bf A} -
\frac{\mu}{c} \mbox{\boldmath$\sigma$}
\times {\bf{E}} \right)^2
+e A_0- \mu \mbox{\boldmath$\sigma$} \cdot {\bf{B}},
\end{equation}
where $\sigma_i$ with $i=1,2,3$ are the Pauli matrices and $\mu = { e\hbar  /(2mc)}$ is a magnetic moment
of the electron. 
This Hamiltonian has $U(1) \otimes SU(2)_{spin}$
gauge symmetry with $U(1)$ gauge potential $A_\nu =(\phi, {\bf A})$
and $SU(2)_{spin}$ gauge potential $b_\nu = (b_0, {\bf b})=(- \mbox{\boldmath$\sigma$} \cdot {\bf{B}},
\mbox{\boldmath$\sigma$}
\times {\bf{E}}) $ \cite{Anandan}. 
 The electric field ${\bf{E}}$ and the magnetic field ${\bf B}$ are measured in the inertial laboratory coordinate
system in which Hamiltonian (\ref{Eq:One}) is written.

{This Hamiltonian includes all gauge interactions related to
our general case, however,
we first focus on the AC effect to concentrate on studying additional spin dependent
forces due to the rotation of the rest frame of the spin and so it is supposed $A_\nu=0$ and $b_0=0$.
Without loss of generality, the electron is supposed to move in a ring of radius $a$ for simplicity.
Then in cylindrical coordinates $(r, \phi, z)$, the Hamiltonian of the electron becomes
$
\label{Eq:OH}
\mathcal{H}=\frac{1}{2 m a^2}\left(-i \hbar \frac{d}{d \phi} - \frac{\mu a}{c} b_{\phi} \right)^2
$, 
where $b_{\phi}={\bf b}\cdot \hat{\phi}$.

The gauge potential ${\bf b}$ helps us to compare two wavefunctions at different space. 
Then the evolution of the wavefunction $\Psi(\phi)$ along the ring 
is described by the equation of parallel transport
\begin{eqnarray}
\label{Eq:SchT}
i\hbar \frac{d}{d\phi} \Psi(\phi) = - \frac{\mu a}{ c}b_\phi \Psi(\phi).
\end{eqnarray}
This equation is a Schr\"odinger-type equation as a function of the azimuthal angle $\phi$, 
instead of time $t$.
Let $\hat{H}(\phi)= -\frac{\mu a}{c} b_\phi$, then an invariant operator
$\hat{I}(\phi)$
satisfies the quantum Liouville-type equation
\begin{eqnarray}
\label{Eq:QL}
i \hbar \frac{\partial}{\partial \phi} \hat{I}(\phi) + [ \hat{I}(\phi), \hat{H}(\phi)]=0.
\end{eqnarray}
Let us suppose the invariant operator $\hat{I}(\phi)$ be known by some technique, then
the exact quantum state of the Schr\"odinger-type equation (\ref{Eq:SchT}) is given by
\cite{Lewis}
\begin{eqnarray}
\label{Eq:WF}
\Psi(\phi) = e^{i \left(\int^\phi_0 \tilde{\Psi}^\dagger_\lambda (\phi') \left( -\frac{1}{\hbar} \hat{H}(\phi')
+ i\frac{\partial}{\partial \phi'} \right)\tilde{\Psi}_\lambda(\phi') d\phi' \right)}\tilde{\Psi}_\lambda(\phi) 
\end{eqnarray}
where $\tilde{\Psi}_\lambda(\phi)$ is an eigenfunction of the invariant operator
$\hat{I}(\phi)$, $\hat{I}(\phi)\tilde{\Psi}_\lambda(\phi) =\lambda \tilde{\Psi}_\lambda(\phi)$.

To study a specific and illustrative model, consider a cylindrically
symmetric electric field ${\bf E}(\phi) = E(\cos{\chi} \hat{\bf r} -\sin{\chi}\hat{\bf z})$,
$\chi$ is the tilt angle with respect to the plane on which the ring lies.
Then $b_\phi$ becomes
$
{ E} \left(\cos{\phi} \sin{\chi} \sigma_1 + \sin{\phi}\sin{\chi} \sigma_2 + \cos{\chi}\sigma_3 \right)
$.
It is easily shown that the invariant operator $\hat{I}(\phi)=\left(\cos{\phi} \sin{\beta} \sigma_1 + \sin{\phi}\sin{\beta} \sigma_2 + \cos{\beta}\sigma_3 \right)$ with $\mu a E/(2\hbar c)=\tan{\beta}/(\cos{\chi}\tan{\beta}-\sin{\chi})$ 
satisfies the quantum Liouville-type equation (\ref{Eq:QL}).

 The eigenvalue equation of $\hat{I}(\phi)$ is solved with
\begin{eqnarray}
\label{Eq:IES}
\tilde{\Psi}_{+}\left( \phi \right)=\left( \begin{array}{c}\cos{\frac{\beta}{2}} \\ e^{i\phi} \sin{\frac{\beta}{2}} \end{array}\right), ~~
\tilde{\Psi}_{-}\left( \phi \right)=\left( \begin{array}{c} -e^{-i\phi}\sin{\frac{\beta}{2}} \\  \cos{\frac{\beta}{2}} \end{array}\right).
\end{eqnarray}
{The angle $\beta$ is the polar angle by which the spin orientation deviates from the $z$ axis.}
The quantum state at $\phi$ is obtained as
\begin{eqnarray}
\label{Eq:TWF}
\Psi_{\pm}(\phi) = e^{i \left(\int^{\phi'}_0 \tilde{\Psi}^*_\pm (\phi') \left(
\frac{\mu a}{\hbar c} b_{\phi'} + i \frac{\partial}{\partial \phi'}\right)\tilde{\Psi}_\pm(\phi')\right)} \tilde{\Psi}^{(\pm)}(\phi),
\end{eqnarray}
for the initial state $\tilde{\Psi}_{\pm}(0)$.
The exponent gives 
{the AC phase} for a cyclic evolution, $\phi'=2\pi$, which is the sum of the dynamical phase
{\begin{eqnarray}
\label{Eq:ACdyn}
\Phi_{\pm}^{1} =\int^{2\pi}_0 \tilde{\Psi}^*_\pm (\phi)  \frac{\mu a}{\hbar c} b_{\phi}\tilde{\Psi}_\pm(\phi) d\phi
= \pm 2 \alpha \cos{(\chi-\beta)} \pi,
\end{eqnarray}
and the spin geometric phase
\begin{eqnarray}
\label{Eq:ACAA}
\Phi_{\pm}^2 = -\int^{2\pi}_0 \tilde{\Psi}^*_\pm (\phi) i\frac{\partial}{\partial \phi} \tilde{\Psi}_\pm(\phi) d\phi = \pm \left( \cos{\beta} -1\right)\pi,
\end{eqnarray}
where $\alpha \equiv\mu E a/(2 \hbar c)$. This geometric phase is obtained for non-adiabatic case
so that this phase is a kind of AA phase.
These $\Phi_{\pm}^1$ and $\Phi_{\pm}^2$ are the same as the dynamical 
and nonadiabatic geometric (AA) phases in Refs. \cite{ Qian}.}


\section{Classical theory and generalized Faraday law}
\label{sec:CF}
One of our main purposes is to understand the spin motive force of an electron in a classical context.
In non-relativistic quantum mechanics classical forces associated with spin could be given by the
generalized Ehrenfest theorem \cite{Yang}
\begin{eqnarray}
\label{Eq:CF}
 m\frac{d {v^i}}{dt} = \langle  \frac{\partial}{\partial t} \left( {p^i} - \frac{\mu}{c} {b^i} \right)  \rangle
+ \frac{i}{\hbar} \langle  \left[ H, {p^i}-\frac{\mu}{c} {b^i}\right] \rangle
\end{eqnarray}
for $v^i=\frac{d}{dt} \langle x^i \rangle=\frac{1}{m}\langle p^i-\frac{\mu}{c}b^i \rangle$, 
where $\langle \hat{O} \rangle $ is the expectation value of an operator $\hat{O}$ in the state $|\Psi\rangle$.

In non-relativistic quantum mechanics spin is described by the covering
group of the rotation group, $SU(2)$ \cite{Sakurai}. The representation of the spin operator 
is given by Pauli matrices representing
non-Abelian properties of the $SU(2)_{spin}$ gauge potential in our case.
The force in Eq. (\ref{Eq:CF}) is determined by average values of observables for 
an wavefunction of the electron. The expectation values of observables become $c$-numbers
even for non-commuting observables so that the force in Eq.  (\ref{Eq:CF}) is
composed of $c$-numbers only. 
Therefore the force in AC effect depends on the expectation values of the electronic spin operators.
Since the average values of the spin operators are $c$-numbers,
the classical spin operators in a proper classical theory are also desirable to have an Abelian nature.

Ryu has derived the spin-motive force in AC effect 
by obtaining the spin-dependent force on a spin
using covariant derivative of the potential term $U=-\frac{\mu}{c} {\bf v}\cdot {\bf b} + \mu b_0$
in the Lagrangian operator $\mathcal{L}=T-U$ \cite{Ryu}. 
This spin-dependent force could be called as a spin-Lorentz force, analogue to the Lorentz force for an electric charge.
The spin Lorentz force is written by
\begin{eqnarray}
\label{Eq:SLF}
{\bf F}_s = e \left( {\bf E}_s + \frac{\bf v}{c} \times {\bf B}_s\right),
\end{eqnarray}
where \\
${\bf E}_s = \frac{\mu}{e}\left(- \mbox{\boldmath$\nabla$}b_0 -\frac{1}{c} \frac{\partial}{\partial t} {\bf b} 
+ i \frac{\mu}{\hbar c}\left[{\bf b}, b_0\right]\right)$ and 
${\bf B}_s = \frac{\mu}{e}\left(\mbox{\boldmath$\nabla$} \times 
{\bf b} - i \frac{\mu}{\hbar c} {\bf b} \times {\bf b}\right)$.
The spin-motive force was defined from this spin Lorentz force as
\begin{eqnarray}
\label{Eq:SMFL}
\mathcal{E} = \frac{1}{e} \oint {\bf F}_s \cdot d {\bf l} 
= -\frac{\mu}{ec} \frac{\partial}{\partial t}\oint{\bf b}\cdot d{\bf l}
\end{eqnarray}
in the analogy to the electromotive force.
Here $\oint d {\bf l}$ is the line integral around the closed path.
We have defined the spin-motive force by the quotient of the line integral 
of the spin Lorentz force divided by the electric
charge $e$ not magnetic charge $\mu$ since it is convenient to give a unified view of a generalized 
Faraday law including both electro- and spin-motive forces.
The average value of this spin-motive force (\ref{Eq:SMFL}) for the electron in the 
ring geometry considered in section \ref{sec:AC} is calculated as 
$\mathcal{E} = -\frac{\partial}{\partial t} _{\pm}\langle \frac{\mu a}{ec} \int b_\phi d \phi \rangle_{\pm}=
-\frac{1}{2 \pi}\frac{d}{dt}\left( \Phi^1_{\pm}\Phi_0\right)$, where
$_{\pm}\langle  \rangle_{\pm} $ represents the expectation value in the wavefunction 
${\Psi}_{\pm} $ in Eq. (\ref{Eq:TWF}).
This spin-motive force contains only the contribution from the dynamical phase, $\Phi^1_{\pm}$.
The spin-Faraday law could not be satisfied since there is no contribution from the spin
geometric phase. This is because the spin Lorentz force (\ref{Eq:SLF}) 
is not a proper classical force since it involves non-commutative operators which must be 
averaged over the quantum state. 

Therefore it is desired to find the classical spin Lorentz force which consists of only commuting variables.
This Abelian nature is achieved by requiring the $z$-direction of the instantaneous coordinate system in the rest 
frame of the spin always coincides with the direction of the spin.
Then the spin operator $\hbar \mbox{\boldmath $\sigma$}/2$ 
and the corresponding magnetic moment operator $\mbox{\boldmath $\mu$}$ 
in this coordinate system has only 
the $z$-component, so that the $SU(2)_{spin}$ gauge field ${\bf b}$
becomes effectively Abelian.
{Note that $\mbox{\boldmath$\mu$}={e} {\bf s}/2mc$}.
In our case where the motion of the spin is represented by the wavefunction (\ref{Eq:TWF}), 
the spin precesses along the $z$ axis of the inertial laboratory coordinate system
so that the $z$-axis of the instantaneous coordinate frame of the spin's rest frame 
also rotates with respect to an inertial laboratory frame.
This means the instantaneous rest frame of the spin is not an inertial reference frame, so
the total time rate of change of the spin in the inertial laboratory frame is given
by
\begin{eqnarray}
\label{Eq:SD}
\frac{d {\bf s}}{d t}={\bf s} \times \left(
\frac{e}{mc}\left({\bf B}-\frac{{\bf v}}{c}\times {\bf E} \right)-
\mbox{\boldmath$\omega$}  \right).
\end{eqnarray}
Here electromagnetic fields ${\bf E}$ and ${\bf B}$ are in the inertial laboratory frame.
$\mbox{\boldmath$\omega$}$
is the angular velocity of the precession of the spin with respect to
the laboratory frame.
The additional term ${\bf s} \times \mbox{\boldmath$\omega$}$ 
is due to the rotation of the rest frame of the spin.

Then the Lagrangian for the spin can be constructed by $\mathcal{L}=T-U$, 
where $T$ is a kinetic energy and
$U$ is a generalized potential energy. $U=-\frac{\mbox{\boldmath$\mu$}}{2} \cdot
\left({\bf B} - \frac{\bf v}{c}\times {\bf E} \right)+ \bf{s} \cdot \mbox{\boldmath$\omega$}$ 
is the corresponding energy of interaction for the mechanical torque in Eq. (\ref{Eq:SD}).
The corresponding Hamiltonian is written by
\begin{eqnarray}
\label{Eq:CC}
 \mathcal{H} = \frac{1}{2m} \left( {\bf p}-\frac{1}{c}\mbox{\boldmath$\mu$} \times {\bf E} + 
 \frac{1}{a} {\bf s}\times {\hat{{\bf r}}} \right)^2
-\mbox{\boldmath$\mu$}\cdot {\bf B},
\end{eqnarray}
where ${\bf v}= a \hat{{\bf r}} \times \mbox{\boldmath$\omega$}$ since the angular velocity
of the precession of the spin is the same as the angular velocity of rotation along the ring 
in this case as seen by (\ref{Eq:TWF}).
Note that the only difference between this Hamiltonian and the Hamiltonian (\ref{Eq:One}) for $A_\mu=0$
is the additional term $\frac{1}{a} {\bf s}\times {\hat{{\bf r}}}$ 
due to the rotation of the rest frame.



Under current situation, we can define
an effective $U(1)$ gauge potential
\begin{eqnarray}
\label{Eq:EffA}
{\bf A}_{eff} = \frac{1}{e}\mbox{\boldmath$\mu$} \times {\bf E} -
\frac{c}{e a} {\bf s} \times {\bf \hat{r}}.
\end{eqnarray}
This is because the spin operator ${\bf s}$ and the magnetic moment operator 
$\mbox{\boldmath{$\mu$}}$ could be considered as
 ordinary commuting vectors since they have only one component in the rotating rest frame
 of the electronic spin \cite{Note}.
 This effective vector potential ${\bf A}_{eff}$ gives the spin-dependent 
 electric and magnetic fields by ${\bf E}_s= -\frac{\partial}{\partial t} {\bf A}_{eff}$
 and ${\bf B}_s= \mbox{\boldmath$\nabla$} \times {\bf A}_{eff}$.
 Then the second term in the ${\bf A}_{eff}$ gives additional spin dependent forces,
 $-\frac{c}{ea} \mbox{\boldmath$\nabla$} \times \left( {\bf s}\times {\bf \hat{r}} \right) $ 
 and $\frac{c}{ea} \frac{\partial}{\partial t} \left( {\bf s}\times {\bf \hat{r}} \right)$. 
 When the ${\bf A}_{eff}$ becomes time dependent, 
 the $-\frac{c}{ea} \mbox{\boldmath$\nabla$} \times \left( {\bf s}\times {\bf \hat{r}} \right) $ 
 does not contribute to the spin-motive force since this term is the same characteristic as 
 the usual magnetic field.  
 Therefore the spin-motive force is given by 
  \begin{eqnarray}
 \label{Eq:SMF1}
\mathcal{E} = \frac{1}{e} \oint {\bf F}_s \cdot d {\bf l}=-\frac{1}{c} \frac{d}{dt}\oint {\bf A}_{eff} \cdot d {\bf l}, 
 \end{eqnarray}
 where ${\bf F}_s=e\left( {\bf E}_s + \frac{\bf v}{c} \times {\bf B}_s\right)$ 
 is the spin Lorentz force. 
 
 {
 Note that $
 \frac{1}{e} \oint \mbox{\boldmath$\mu$} \times {\bf E} \cdot d {\bf l} 
$ gives the dynamical phase (\ref{Eq:ACdyn}) times $\Phi_0$.
 And $\oint \left( -
\frac{c}{e a} {\bf s} \times {\bf \hat{r}}\right) =\Phi_0 \cos{\beta} \pi $.
This gives the same contribution to motive force as the AA phase (\ref{Eq:ACAA})
since the difference between this result and the AA phase is just a constant $-\pi$ except 
another overall constant $\Phi_0$.}
 This implies the effect of the spin geometric phase on
 the motive force could be explained by the additional spin Lorentz
 force generated by the rotation of the coordinate system.
 Therefore the spin-motive force is represented as $-\frac{1}{c}\frac{d}{dt} (\Phi_{AC} \Phi_0)$,
 where $\Phi_{AC}$ is the AC phase.
 This implies the spin Faraday law is satisfied for the AC effect in general.
 }
 
 Next we will consider the spin Faraday law for the scalar AC effect.
 In the scalar AC effect the spin-motive force is generated by a time-dependent magnetic field.
 In Stern's geometry \cite{Stern} , ${\bf B}= -{\bf B}_\phi(t) \sin{\phi}\hat{x} + {\bf B}_\phi(t) \cos{\phi}\hat{y}+
 B_z\hat{z}$ and ${\bf E}={\bf 0}$, the Hamiltonian (\ref{Eq:One}) becomes
 \begin{eqnarray}
 \label{Eq:SACH}
 \mathcal{H} =\frac{1}{2 m a^2}\left(-i \hbar \frac{d}{d \phi} - \frac{ e B_z \pi a^3}{2c} \right)^2 -\mu {\bf B}\cdot \mbox{\boldmath{$\sigma$}}.
 \end{eqnarray}  
 Let $H_0=\frac{1}{2 m a^2}\left(-i \hbar \frac{d}{d \phi} - \frac{ e B_z \pi a^3}{2c} \right)^2$
 and a wavefunction $\Psi_0$ be the eigenfunction of $H_0$, i.e., $H_0\Psi_0 = E_0 \Psi_0$
 with $E_0=\frac{1}{2ma^2}\left(n - \frac{ e B_z \pi a^3}{2c} \right)^2$, where $n$ is an integer.
 
 If we write the wavefunction $\Psi$ which satisfies the time-dependent Schr\"odinger equation
 given by the Hamiltonian (\ref{Eq:SACH})
 as the product of $\Psi_0$ and $\Psi_1$, then the time-dependent Schr\"odinger equation for $\Psi_1$
 is acquired as
$ i\hbar\frac{\partial}{\partial t} \Psi_1= -\mu {\bf B}\cdot \mbox{\boldmath{$\sigma$}} \Psi_1.$
The quantum Liouville equation with $\mathcal{H}_1=-\mu {\bf B}\cdot \mbox{\boldmath{$\sigma$}}$ 
is satisfied by the invariant operator,
$
\hat{I}= - \sin{\phi} \sin{\chi} \sigma_1 + \cos{\phi}\sin{\chi} \sigma_2 + \cos{\chi}\sigma_3
$.
The eigenvalue equation $\hat{I} \tilde{\Psi}_1^\lambda = \lambda \tilde{\Psi}_1^\lambda$ is 
solved with the eigenfunctions
\begin{eqnarray}
\tilde{\Psi}_1^+ = \left( \begin{array}{c} e^{-i\phi} \cos{\frac{\chi}{2}} \\ i \sin{\frac{\chi}{2}} \end{array}\right),
\tilde{\Psi}_1^- = \left( \begin{array}{c}- e^{-i\phi} \sin{\frac{\chi}{2}} \\ i \cos{\frac{\chi}{2}} \end{array}\right),
\end{eqnarray}
where $\tan{\chi} = \mu B_\phi /(\hbar \omega + \mu B_z)$ and $\omega=d\phi/dt$ 
is the angular velocity of the particle encircling the ring. 
The geometric phase accumulated by the particle encircling 
the ring is obtained as
$
\phi^{\pm}_g = -\int (\tilde{\Psi}_1^{\pm})^\dagger i \frac{\partial}{\partial t} 
({\tilde{\Psi}_1^{\pm}}) = \pi (1\pm \cos{\chi})$,
which is half the solid angle subtended by the cone swept by the spin magnetic moment.
{
It is observed when $\hbar \omega \ll \mu B_z$, the adiabatic condition holds and the angle $\chi$ between the 
$z$ axis and the spin magnetic moment becomes approximately equal to 
the angle between the $z$ axis and the magnetic filed which
is the same as $\alpha$ in Ref. \cite{Stern}.

{The dynamical phase $\int_0^\tau (\tilde{\Psi}_1^{\pm})^\dagger 
( - \mu {\bf B} \cdot \mbox{\boldmath{$\sigma$}})({\tilde{\Psi}_1^{\pm}}) dt$ does not contribute to
the motive force since the spin-dependent force $\mu \mbox{\boldmath{$\nabla$}}({\bf B} \cdot \mbox{\boldmath{$\sigma$}})$
related to the dynamical phase is always conservative, that is, 
$\oint \mu \mbox{\boldmath{$\nabla$}}({\bf B} \cdot \mbox{\boldmath{$\sigma$}}) \cdot d{\bf l}=0$.
Where $\tau$ is the time spent for one cyclic motion.}
When the magnetic filed is time dependent the geometric phase becomes time-dependent,
and the particle is subject to a motive force 
$\mathcal{E}^{\pm}=-\frac{d( \phi_g^{\pm} \Phi_0)}{d t}$ in analogy with
Faraday law.

In this case the spin is also precessing under the magnetic field ${\bf B}$ 
so that the rest frame of the spin is 
rotating with respect to the laboratory frame. Therefore a classical Lagrangian for this spin has
the additional term $-{\bf v} \cdot ({\bf s} \cdot \hat{r})/a$ originated from the rotation of the electron's 
rest frame.
 Then the motive force derived by the classical spin-dependent force becomes
\begin{eqnarray}
\mathcal{E} = \frac{1}{c}\frac{d}{dt}\int\left( \frac{c}{ea} {\bf s}\times \hat{r} \right)\cdot
d {\bf l} = -\frac{d}{dt}\left({\Phi_g^+ \Phi_0}\right),
\end{eqnarray}
since the only spin Lorentz force contributing to spin motive force is 
$\frac{1}{a} \frac{\partial}{\partial t} \left( {\bf s}\times {\bf \hat{r}} \right)$.
Therefore the motive force derived by the classical force satisfies the spin Faraday law.

The motive force is easily generalized for the case of the Hamiltonian (\ref{Eq:One}) 
in which both $U(1)$ electromagnetic and $SU(2)_{spin}$ interactions coexist. 
The AB and SAB phases could be understood on equal footing as the phase accumulated
to a wavefunction through a cyclic evolution 
by ${\left( \frac{e}{\hbar c} \oint A_\nu dx_\nu\right)}$, where $\nu=0,1,2,3$ and
$A_{\nu}=(\phi, {\bf A} )$ is an $U(1)$ electromagnetic four vector potential.
Using the $SU(2)_{spin}$ gauge field we could also describe the AC phase and
the SAC phase together.
The total phase acquired by the electron in cyclic motion is given by an exponent of 
the phase factor $\exp{\left(\frac{\mu}{\hbar c} \oint b_\nu dx_\nu\right)}$ for
the special initial state, which returns to the initial one after a cyclic evolution
apart from a phase factor.
The total phase, however, is not the simple sum of the AC phase and the SAC phase
since the $SU(2)_{spin}$ gauge potentials do not commute.

  The phase accumulated to a wavefunction in the general case
is obtained by the parallel transporter
\begin{eqnarray}
\exp{\left( \frac{e}{\hbar c} \int_C A_\nu dx^\nu\right)}
\mathcal{P}\exp{\left( \frac{\mu}{\hbar c} \int_C b_\nu d x^\nu\right)}.
\end{eqnarray}
{Since the electric potential $\phi$
is given by a conservative force $-\mbox{\boldmath{$\nabla$}} \phi$, only the AB phase
given by the magnetic vector potential
contributes to motive forces. The second path integral is path-ordered since
it contains non commuting operators. This path integral gives AC and scalar AC phase 
under a cyclic evolution with a special wavefunction which returns its original form apart from 
the overall phase factor. The total dynamical phase becomes the simple sum of the dynamical
phase in the AC effect and that in the scalar AC effect.
This is because the path-ordered integral could be performed by invariant operator method.
In Eq. (\ref{Eq:WF}) the dynamical phase
is given by the integral of the averaged operators which becomes effectively commutative.
The nonadiabatic geometric phase, however, is determined by the geometry of the precession of the 
spin determined by the combined effects of 
both $SU(2)_{spin}$ gauge fields, ${\bf b}$ and $b_0$. 
That is, the precession angle of the spin is not the simple sum of
the precession angles for AC and scalar AC effects. 
Therefore the spin geometric phase (AA phase) does not become just the simple sum of
 each AA phases in the AC and SAC effects.}

{The dynamical phase in the AC effect contributes to the motive force as shown in section \ref{sec:AC}.
The dynamical phase in the SAC effect, however, does not contribute to motive forces
as we have seen in the above Stern's example.
Therefore the phases contributing to motive forces are the AB phase and
the dynamical phase of the AC phase and the spin geometric phase which is given by
the combined effect of the scalar and vector $SU(2)_{spin}$ gauge fields.
The existence of the geometric phase implies the precession of the spin so that the instantaneous
rest frame of the spin is rotating. This rotation of the rest frame gives an additional 
effective $U(1)$ gauge potential $- \frac{c}{ea} {\bf s}\times \hat{r}$ in the
classical Hamiltonian as shown.
Hence in the rotating rest frame of the electron spin, 
the effective $U(1)$ gauge potential under $SU(2)_{spin}$ gauge interaction
becomes the same as ${\bf A}_{eff}$ of Eq. (\ref{Eq:EffA}).

The spin dependent forces that produce generalized motive force are 
$-\frac{e}{c}\frac{\partial}{\partial t} {\bf A}$
 and $-\frac{e}{c}\frac{\partial}{\partial t} {\bf A}_{eff}$,
 where ${\bf A}$ is the magnetic vector potential. 
Therefore when these phases vary in time, they generates the generalized motive force as
\begin{eqnarray}
\mathcal{E} = \frac{1}{e} \oint {\bf F}\cdot d {\bf l} 
=-\frac{d}{dt} \oint \left( {\bf A} +{\bf A}_{eff} \right). 
\end{eqnarray}
This generalized motive force is easily calculated and could be represented as
$-\frac{1}{2\pi}\frac{d}{dt} \left(\Phi_{AB}+ \Phi_{dyn,AC}+ \Phi_{geo}\right)\Phi_0$. 
Here $\Phi_{AB}$, $\Phi_{dyn,AC}$, and $\Phi_{geo}$ are the AB phase,
the dynamical phase in the AC effect, and the spin geometric phase respectively.
Let $\Phi_T=\Phi_{AB}+ \Phi_{dyn,AC}+ \Phi_{geo}$, then the same form with the ordinary
Faraday law is acquired for the generalized Faraday law including both
the electro- and the spin-motive forces as
$
\mathcal{E}= -\frac{1}{2\pi}\frac{d}{dt} \left(\Phi_T \Phi_0\right)
$.}


\section{Conclusion and Discussion}
\label{sec:DIS}

In summary we have obtained the generalized Faraday law 
in a unified manner using the classical spin Lorentz forces in the rotating rest frame of
the electron spin. The spin is
intrinsically quantum object, however we have shown that the spin
can be successfully described by a classical magnetic moment when
the rotation of the rest frame is correctly considered. In this
picture the effect on the spin-motive force of the time varying 
spin geometric phase is understood as the same as that of the
new spin-dependent classical force generated by the spin precession. 

The motive force in Ref. \cite{Stern} has an amplitude of
$10^{-7}V$ in adiabatic condition. In the nonadiabatic case 
the magnitude of $\hbar \omega$ in $\tan{\chi}$
is approximately $10^{-23} J$ and $\mu B_z \approx 9.27\times 10^{-24} J$ for
$B_z=1 T$ so that 
the effect of time-dependent non-adiabatic geometric phase on
the generalized motive force could also be observed.
   The model Hamiltonian of Refs. \cite{Barnes, Yang1} have similarities
to that of Ref. \cite{Stern}, since the interaction between the internal exchange field
and the spin is the Zeeman-type.
With our picture the physical origin of the generalized Faraday law 
mentioned in Refs. \cite{Barnes, Yang} is satisfied could be clearly understood.

\section*{Acknowledgements}
 This work was supported by the Korea Research Foundation Grant(KRF-2008-331-C00073), 
 a research grant from Seoul Women's University(2009).
 We gratefully acknowledge KIAS members and Dr. Sam Young Cho for helpful discussions.


\begin{thebibliography}{99}
\bibitem{Note} For the spin with eigenvalue $-$ in Eq. (\ref{Eq:IES})
  an effective $U(1)$ gauge potential 
 becomes $- {\bf A}_{eff}$ which is just the negative of the effective $U(1)$ gauge potential
 for the spin with the $+$ eigenvalue. So we focus on the spin with $+$ in the followings.
\end{thebibliography}

\end{document}